# Heat and Hostility: How Substrate Temperature Shapes Bacterial Deposition Patterns and Pathogenesis in Evaporating Droplets

**Amey Nitin Agharkar[1], Anmol Singh[2], Kush Kumar Dewangan[3], Dipshikha Chakravortty[2,4]* and Saptarshi Basu[1,3]***

[1] Interdisciplinary Centre for Energy Research (ICER), Indian Institute of Science

[2] Department of Microbiology & Cell Biology, Indian Institute of Science

[3] Department of Mechanical Engineering, Indian Institute of Science.

[4]Adjunct Faculty, School of Biology, Indian Institute of Science Education and Research, Thiruvananthapuram

Conflict of Interest – The authors have declared that no conflict of interest exists.

*Corresponding author

**Address correspondence to**: Prof. Saptarshi Basu, 406, Department of Mechanical Engineering, Indian Institute of Science (IISc), Bangalore 560012, India, Telephone number: +91 80 22933367, Email: sbasu@iisc.ac.in

Prof. Dipshikha Chakravortty, SA-10, Department of Microbiology and Cell Biology, Indian Institute of Science (IISc), Bangalore 560012, India, Telephone number: +91 80 22932842, Email: dipa@iisc.ac.in

# Abstract

**Hypothesis**

Droplets ejected from the host can directly settle on a substrate as fomite. In industrial environments, especially the food processing industries, the components maintained at specific temperatures can act as a substrate, leading to the fomite mode of infection. We hypothesize that substrate temperature influences the desiccation dynamics, bacterial deposition patterns, and bacterial viability and infectivity.

**Experiments**

We conducted a novel study on the desiccation behaviour of bacteria-laden droplets on hydrophilic substrates at different temperatures, an area rarely explored. Such studies have been rarely attempted. We analysed bacterial deposition patterns, mass transport dynamics, and viability across various base fluids used in food industry, such as Milli-Q water, LB media, and meat extract. Thermal imaging, confocal microscopy, scanning electron microscopy, atomic force microscopy, and optical profilometry characterized pattern formations, while bacterial viability and infectivity were assessed post-desiccation

**Findings**

Our results indicate that substrate temperature significantly affects bacterial deposition and viability. With Milli-Q water, lower temperatures resulted in ring-like deposits, while higher temperatures led to thinner rings with inner deposits due to Marangoni convection. Radial velocities at 50°C were an order of magnitude higher than 25°C. For LB media, dendritic patterns varied with temperature, whereas meat extract patterns remained unchanged. At 60°C, bacterial surface area was significantly reduced compared to 25°C while maintaining a constant aspect ratio. Higher temperatures reduced bacterial viability in precipitates, but bacterial abstractabstractinfectivity remained nearly unchanged across all base fluids. These findings highlight potential fomite-based infection risks from heated surfaces, particularly in industrial settings.

**Keywords:** Sessile, Hydrophilic, Temperature, Evaporation, Bacteria, Viability, Virulence

# 1. <u>Introduction</u>

Droplets ejected from a host containing pathogens can either desiccate in air and form nuclei, desiccate for a specific time and then fall on a surface, or directly settle on a surface and evaporate as a sessile droplet [1-10]. Sessile droplets evaporate and form pathogen depositions on the surface, promoting the fomite mode of infection. Contaminated surfaces, known as fomites, are critical in transmitting microbial infections across community, healthcare, and industrial environments [11],[12]. Fomites act as reservoirs for bacterial pathogens, contributing to outbreaks of both community-acquired and hospital-associated (nosocomial) infections.

In the food industry, due to lack of sanitization, surfaces such as conveyor belts, cutting tools, and packaging materials can harbour pathogens like Listeria monocytogenes, *Salmonella* spp., and Escherichia coli, causing large-scale foodborne outbreaks [13],[14]. Microbial contamination of surfaces in pharmaceutical manufacturing presents significant health risks to consumers [15]. Key contact points for fomite-mediated transmission among workers include high-touch surfaces in industrial workspaces, such as machinery controls, handrails, and shared equipment [16]. In industrial settings, components are maintained at specific temperatures. Consequently, if an infected droplet lands on an isothermal surface and evaporates, assessing the potential for fomite-based transmission becomes essential. Temperature significantly influences the viability and infectivity of pathogens by regulating the desiccation dynamics [17]. Therefore, understanding the influence of substrate temperature on bacterial survival and virulence is vital for evaluating the risk of fomite-based infection, yet such studies have been rarely conducted.

The bacteria *Salmonella* is one of the leading causes of foodborne illness globally and a key contributor to acute gastroenteritis. Each year, it is estimated to cause over 300,000 deaths, primarily in underdeveloped countries [18]. The most common serovars, *Salmonella enterica* serovar Typhimurium (STM) and Enteritidis are major contributors to nontyphoidal salmonellosis infections linked to human diseases such as acute gastroenteritis [19]. Eliminating *Salmonella* from the food chain remains challenging due to its resistance to environmental stressors. Its diversity and adaptability make detection in food and food-related environments difficult, underscoring the importance of verifying process controls.

Exceptional adaptability to extreme conditions is demonstrated by *Salmonella* as it survives in harsh environments such as extracellular pH levels ranging from 3.99 to 9.5, salt concentrations up to 4% NaCl, and temperatures between 2°C and 54°C. It regulates its stress responses through transcriptional regulators (e.g., *SoxS/SoxR, OxyR, Fur, RamA, RamR, MarA,* and *MarR*), phospho-relay-based two-component systems (e.g., *BaeRS, CpxRA, OppR-EnvZ, PhoPQ, and PmrAB [BasRS]*), and alternative sigma factors (e.g., σS, σE, and σH) [20]. These regulatory mechanisms activate gene expression linked to specific stress responses, with some genes overlapping multiple stress pathways [21].

*Salmonella* infections primarily spread through consuming contaminated food, including raw or undercooked meat, poultry products, eggs, unpasteurized milk, other dairy products, and raw fruits and vegetables [22]. Research indicates that *Salmonella* survival at 70°C significantly increases after six days of incubation in peanut oil or exposure to the sub-lethal heat treatment at 45°C for three minutes [23]. Additionally, *Salmonella* has been shown to exhibit enhanced heat resistance and survivability in low-moisture foods [24],[25].

In this study, we addressed the fundamental question of bacterial survival and infectivity in sessile droplets on hydrophilic glass substrates maintained at temperatures ranging from 25°C to 70°C, using physiologically relevant fluids such as LB media and meat extract. Our experiments reveal that higher substrate temperatures accelerate evaporation and intensify thermal Marangoni convection, leading to significant changes in bacterial deposition patterns and cell size parameters—a novel conceptual advance in colloid and interface science that contrasts with earlier studies focused solely on nanoparticle-based droplet studies. Although the overall bacterial survival decreases with increasing temperature, Salmonella maintains its virulence up to 60°C substrate temperature, posing alarming food safety concerns. Our work uniquely bridges colloidal dynamics with microbiological assessments, offering new insights into how thermal and interfacial forces govern bacterial behaviour under stress. The interdisciplinary approach of the study not only expands the current framework of colloid and interface science but provides a novel perspective on pathogen survival mechanisms that can inform improved industrial practices and risk management in food processing.

# 2. Experimental Methods

## 2.1 Sample Preparation

### 2.1.1 Bacteria, Plasmid, and growth conditions

All the experiments used the *Salmonella enterica* Serovar Typhimurium 14028s (STM WT) strain. The bacterial strain was cultured in Luria broth (LB-Himedia) with constant shaking (170 rpm) at 37°C orbital shakers. Strains were transformed with a pFPV-cherry plasmid for immunofluorescence assays; for the culture STM containing pFPV-cherry plasmid, 50µg/ml ampicillin antibiotic was used.

### 2.1.2 Base fluid preparation

Mill-Q water, LB (1% Trypton, 0.5 % Yeast extract, 0.5% NaCl) (HIMEDIA), and 1% meat extract (M807 HIMEDIA) were used as base fluid. All fluids were sterilized and autoclaved at 121°C and a pressure of 15 psi.

### 2.1.3 Bacterial sample preparation

STM WT or STM WT containing pFPV-mcherry plasmid (STM WT RFP) was inoculated in LB media from a freshly streaked LB agar plate. The overnight culture was washed twice with autoclaved Milli-Q water and adjusted to an optical density (OD) corresponding to $1\times10^8$ bacteria per ml in Milli-Q, LB broth, and meat extract.

## 2.2 Experimental setup

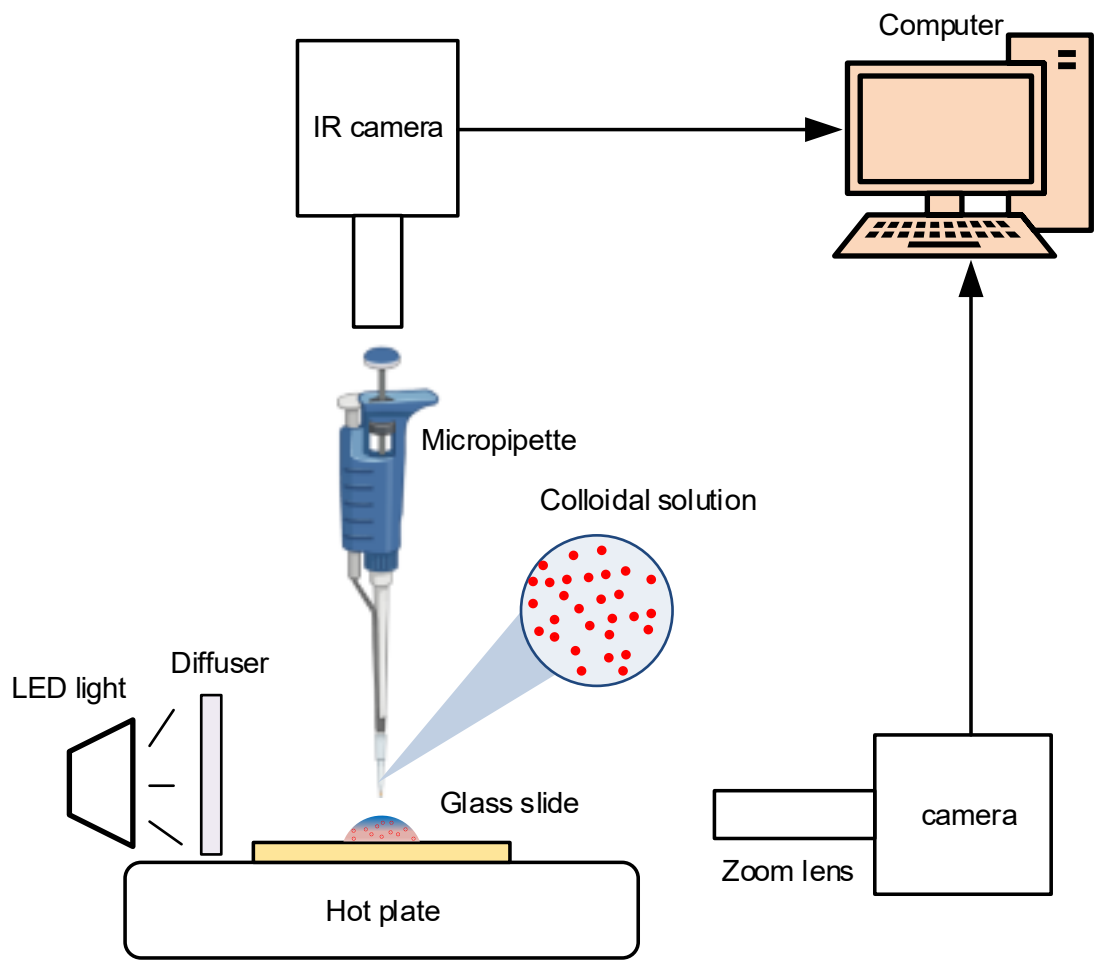

**Figure 1: Schematic of the experimental setup for evaporation studies and thermal imaging of the sessile droplet on a hydrophilic substrate maintained at various temperatures on a heater plate.**

Figure 1 illustrates the experimental setup used in this study. The ambient temperature was maintained at 25°C ± 1°C, and the relative humidity (RH) was kept at 50% ± 2%. Temperature and humidity measurements were taken using a TSP-01 sensor from Thorlabs for validation. A commercial DSLR camera (Nikon D5600) equipped with a Navitar zoom lens (2X lens × 4.5X tube) was used to capture the lifetime of a sessile droplet at 30 frames per second, with a spatial resolution of 1pixel/μm. To ensure uniform illumination, a diffuser plate was placed between a 5W LED light source and the sessile droplet. A 1 μL droplet was deposited onto a hydrophilic glass slide, which was placed on a heater plate. The heater plate assembly of Delta Power Controls (Mains: 230 V,15 A, Output: 50 V, 3A) was utilized for the experiments maintaining a constant plate temperature. The outer body of the plate is made up of stainless steel, heating coil of nichrome wire and heating chamber body made up of copper. Thyristor Proportional–integral–derivative (PID) temperature controller system was used to attain a particular set temperature, which was cross-verified with a K-type Thermocouple. The heater plate was maintained at six set experimental temperature points: 25ºC, 37ºC, 45ºC, 55ºC, 60 ºC and 70 ºC. Glass side (Bluestar, Polar industries :25 mm × 25 mm × 1.35 mm thickness) is placed on the heater plate before it starts heating, and it is allowed to achieve steady state temperature for 3 minutes before drop casting (refer Supplementary Material Section S1). Prior to use, the glass slides were cleaned by sonication in a propan-2-ol bath for 2-4 minutes, followed by wiping with Kimwipes. The side view of the drying sessile droplet was recorded using the camera setup. Given that the Capillary number ($Ca = \mu u/\gamma \sim 10^{-8}$) and Bond number ($Bo = (\rho g R_c\, h/\gamma) \sim 10^{-2}$) are consistent with the spherical cap model, the droplet volume was estimated using the equation: $V_s = (\pi h/6)(3R_c^2 + h^2)$ where $R_c$ is the contact radius, and h is the drop height at centre [26]. The temporal changes in the droplet's height and radius during evaporation were tracked using the "Analyze Particles" plugin in ImageJ. Thermal imaging of the evaporating droplet was performed using the IR Camera (FLIR SC 5200) with 3X lens assembly at 50 fps for 25ºC and 37ºC cases, 100 fps for 55ºC case and at 150 fps for 45ºC, 55ºC,60 ºC and 70 ºC cases. The results were analysed with Altair software.

## 2.3 Precipitate characterization and analysis

### 2.3.1 Confocal microscopy sample preparation & methods

Confocal samples were prepared using the overnight grown stationary phase culture *Salmonella Typhimurium* Wild type (STM WT) Red fluorescent protein (RFP) culture. The bacteria were pelleted and washed twice using PBS before resuspending. The bacterial concentration was maintained at $10^8$ CFU/mL using the OD 600 technique. Pelleted bacteria are resuspended in 500 µL Milli-Q water, Meat extract and LB media fluids. Zeiss LSM 880 NLO upright multi-photon confocal microscope at 10X magnification is used to image the evaporated precipitate samples. We examined the bacterial deposition within the precipitates using the maximum intensity projection images created with ZEN Black software. (Carl Zeiss).

### 2.3.2 Scanning Electron Microscopy (SEM) of the samples

The same protocol is applied for preparing samples for scanning electron microscopy (SEM), with the bacterial concentration maintained at $10^8$ CFU/mL. The base fluids used in the experiments included Milli-Q water, meat extract, and LB media. For SEM imaging, the precipitate samples on glass slides were gold-coated for 1 minute using Bal-Tec SCD005 sputter coater (220 V, 100VA). Backscattered electron (BSE) imaging was performed using the JEOL-SEM IT 300 microscopy system, which operates with a high probe current of 30 nA, an acceleration voltage of 15 kV, and a working distance of 12 mm.

### 2.3.2 Live cell confocal imaging

The sample preparation is similar to the confocal imaging. Andor Dragonfly 400 confocal microscopy live imaging system was utilized to record evaporation videos of $10^8$ CFU/mL concentration of STM WT in Milli-Q water droplet at droplet edges. The plane of velocity measurement was approximately at a height of 25 µm above the substrate. Average velocities at the two cases of the substrate temperature and the available device RH are analysed, low evaporation rate case: 25°C substrate temperature & 90 ± 5% RH and high evaporation case of 50°C substrate temperature & 54 ± 3% RH. The low evaporation rate case videos were recorded at an exposure time of 130 ms and high evaporation rate case videos at 20 ms respectively. The videos were analysed using PIVLab a MATLAB plugin [27]. Velocities are analysed in four regions of 200 μm × 150 μm starting from the droplet edge towards centre i.e. Region1,2,3, and 4.

### 2.3.4 Atomic Force Microscopy (AFM) imaging of the samples

The samples are prepared in Milli-Q water as base fluid and the bacterial concentration was maintained at $10^7$ CFU/mL. Precipitates of 25°C and 60°C substrate temperature are used for imaging. The Park NX10 atomic force microscope is used to capture the images. For imaging, a CONTSCR probe with a stiffness of 0.2 N/m is employed. Nanoindentation is carried out at a speed of 0.3 μm/s in both the upward and downward directions, without holding the probe in place. The images were analysed using ImageJ.

### 2.3.5 Optical profilometry of the samples

Precipitates of STM WT $10^8$ CFU/mL in Milli-Q water as base fluid are taken for optical profilometry to measure the surface profile. Taylor Hobson Optical Profiler at 20× with Talysurf CCI (Coherence Correlation Interferometry) Lite Systems at optical resolution of 0.4-0.6 μm was utilized for measurements. The precipitate samples were gold coated before measurements.

## 2.3.6 In vitro bacterial viability assessment

The sessile precipitates at different substrate temperature were reconstituted in 10μL of sterile, autoclaved PBS. 990μL of sterile PBS was added to the 10μL retrieved droplet precipitate, and 100μL of it was plated on SS agar plates. For CFU numbers were enumerated before and after heat treatment at different ranges of temperatures: 25°C, 37°C, 45°C, 55°C, 60°C, and 70°C. The Plating was done on *Salmonella-shigella* (SS) agar using the spread plate method and incubated at 37°C overnight.

## 2.3.7 Intracellular infection studies

STM WT samples were prepared after giving heat treatment and collected in PBS and immediately started the infection. More bacterial drops are collected at higher temperatures to keep an equal number of bacterial cells. Plating was always done in the pre-inoculum of the infection samples to normalize the data.

$1.5 \times 10^5$ RAW264.7 macrophages were seeded in each well of 24 well plates and incubated for 7-8 hours at 37°C, 5% CO2. Bacterial attachment to host cells was enhanced by centrifugation at 600 rpm for 6 minutes, then incubating the infected cells for 25 minutes. Next, cells were washed with PBS and treated with gentamicin (100 μg/ml in complete media Dulbecco's Modified Eagle Medium (DMEM)+10 percent Fetal bovine serum (FBS)) for 1 hour to remove all the extracellular bacteria and then maintained with 25μg/ml of gentamicin in complete media for the rest of the experiment. Firstly, the cells were lysed with 0.1% triton-X 100 in PBS at 2-hour time point. The lysate was plated on SS agar, and the corresponding CFU at 2 hours was determined. Percent phagocytosis was determined by using a formula-

[ No. of CFU at 2 hours]/ [no. of pre-inoculum CFU] ×100

To determine fold proliferation, the cells were lysed with 0.1% triton-X 100 in PBS at 16 hours post-infection. The lysate was plated on SS agar, and the corresponding CFU was determined at 2 and 16 hours. Fold proliferation was determined by using the formula:

[ CFU at 16 hours]/ [ CFU at 2 hours]

### 2.3.6 Statistical analysis for viability and infection

Statistical analyses were performed with GraphPad Prism software. The student's t-test was performed for survival assay. Two-way ANOVA with Tukey's post hoc test was used for grouped data for the infection experiment. The results are expressed as mean ± SD from three independent experiments (N ≥ 3). p values for each experiment are described in figure legends (refer Supplementary Material Section S2).

# 3. Results and Discussion

## 3.1 Evaporation of sessile droplet on heated substrate

An infected droplet can fall on the substrate and evaporate as a sessile droplet, forming fomites. In industrial environments, especially food processing industries, the components are maintained at different temperatures, and the fluid goes through sophisticated processes. Therefore, such components can offer as a substrate for the infected droplet, and the droplet can evaporate and infect through the fomite form of infection. Building on that, for doing in-depth research on the effect of substrate temperature on bacterial survival and infectivity, we studied the sessile droplet evaporation dynamics, flow inside the droplet, droplet surface interface temperature distribution, bacterial deposition patterns, and its aerial size distribution.

The droplet percentage volume reduction is dependent on physical conditions/parameters and material properties like evaporation flux, drop temperature, ambient temperature, relative humidity (RH), vapor concentration field (vapor pressure), flow field inside and outside the drop, thermal conductivity and diffusivity, mass diffusivity, solute properties to name a few. Moreover, it is independent of the initial droplet diameter [28]. For a sessile droplet, all these properties are governed by the substrate properties, as the substrate properties are very crucial for the evaporation of the sessile droplet [29],[30],[31].

Firstly, the evaporation dynamics of bacteria-laden droplets, with Milli-Q water as the base fluid, were studied to deconstruct the effect of substrate temperature. The evaporation study for the substrate temperature of 25°C, 37°C, 45°C, 55°C, 60 °C and 70 °C is performed and the normalized evaporation parameters are plotted.

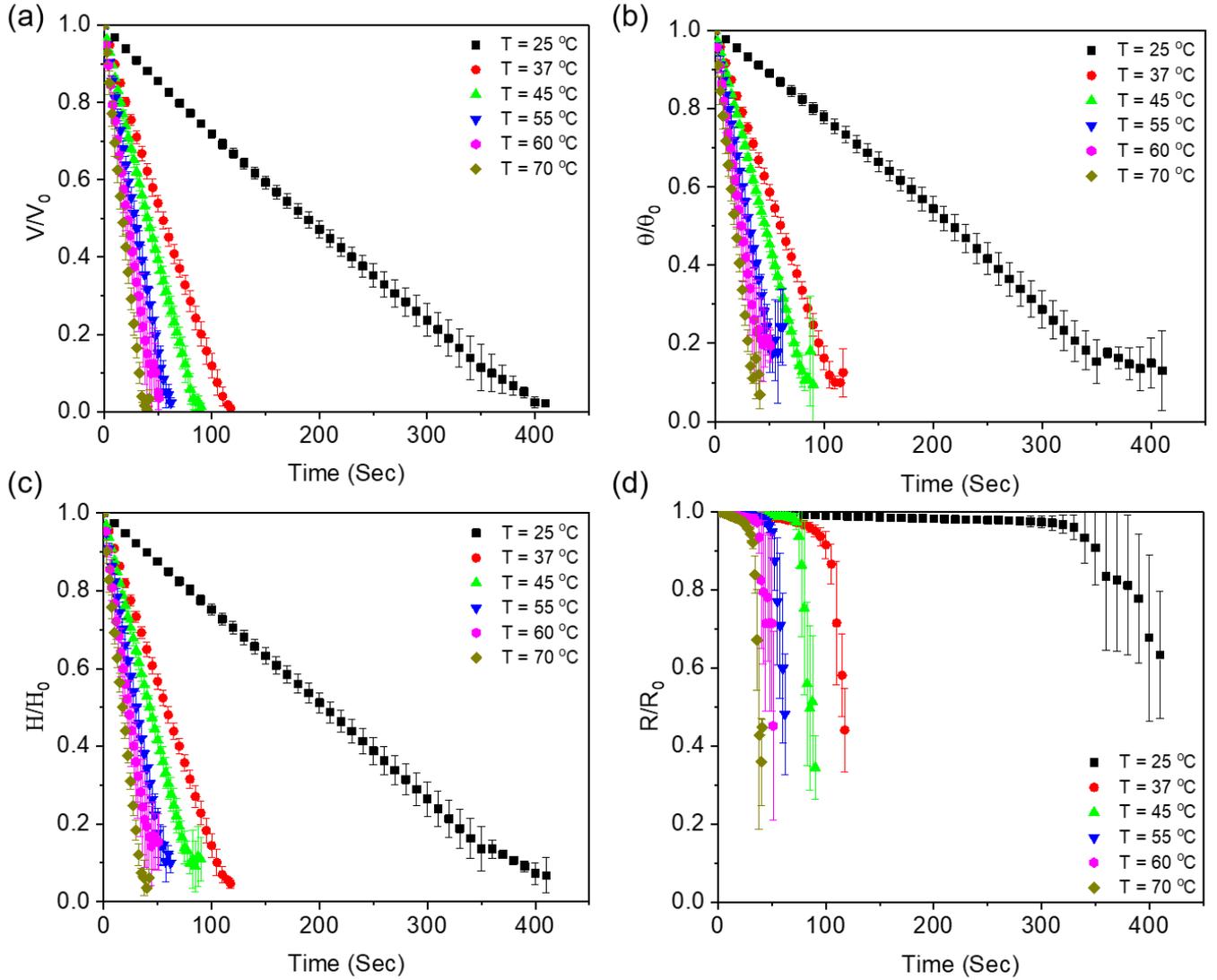

**Figure 2: Normalized droplet parameters versus Time (s) for STM WT in Milli-Q water sessile droplet on substrate at 25°C, 37°C, 45°C, 55°C, 60 °C and 70 °C temperatures (a) Volume ratio $V/V_0$ variation (b) Contact angle $\theta/\theta_0$ variation (c) Height $H/H_0$ variation (d) Wetted Radius $R/R_0$ variation. $V_0$, $\theta_0$, $H_0$, and $R_0$ are the initial volume, initial contact angle, initial height and initial wetted radius of the sessile droplet respectively.**

Figure 2 represents the normalized parameters of sessile droplet evaporation for various substrate temperatures. Figure 2a represents the droplet volume ratios $V/V_0$ (V is the instantaneous volume and $V_0$ is the initial droplet volume) with respect to time in seconds of the evaporating droplet. Similarly Figure 2b, 2c and 2d represents the droplet contact angle ratios $\theta/\theta_0$ ($\theta$ is the instantaneous contact angle and $\theta_0$ is the initial droplet contact angle), droplet height ratios $H/H_0$ (H is the instantaneous height and $H_0$ is the initial droplet height), and wetted droplet radius ratios $R/R_0$ (R is the instantaneous radius and $R_0$ is the initial droplet wetted radius) respectively. The evaporation time of the droplet decreases with increase in the substrate temperature. The droplet evaporating on a substrate at 25°C takes approximately t ~ 400-425 s, while at 70 °C substrate temperature it takes approximately t ~ 35-40 s. By looking at how the drop volume changes over time (refer Figure 2a), we can see that the volume decreases at a steady, linear rate for sessile droplets on a heated substrate [30],[32].

Sessile droplets evaporate in constant contact radius (CCR), constant contact angle (CCA) and mixed mode [33],[34],[35]. In CCR mode, the three-phase contact line remains pinned, so the contact radius remains fixed. As the droplet evaporates, its volume decreases, causing its height and contact angle to shrink. In contrast, CCA mode keeps the contact angle fixed while the contact radius decreases. Drops with acute contact angles typically evaporate in CCR mode for most of the process, transitioning to CCA mode only at the final stage, when the drop flattens into a thin film and retracts. The droplet on the glass slide starts with an initial contact angle of about 20°-45°. As commonly observed on glass surfaces, evaporation mainly occurs in the constant contact radius (CCR) mode [36]. For bacteria-laden Milli-Q water droplet on a heated substrate, the evaporation process occurs majorly in CCR mode (refer Figure 2d) as the wetted radius remains constant for most of the evaporation time. The transition of CCR mode to CCA mode happens towards the end of the evaporation process when the contact line starts depinning.

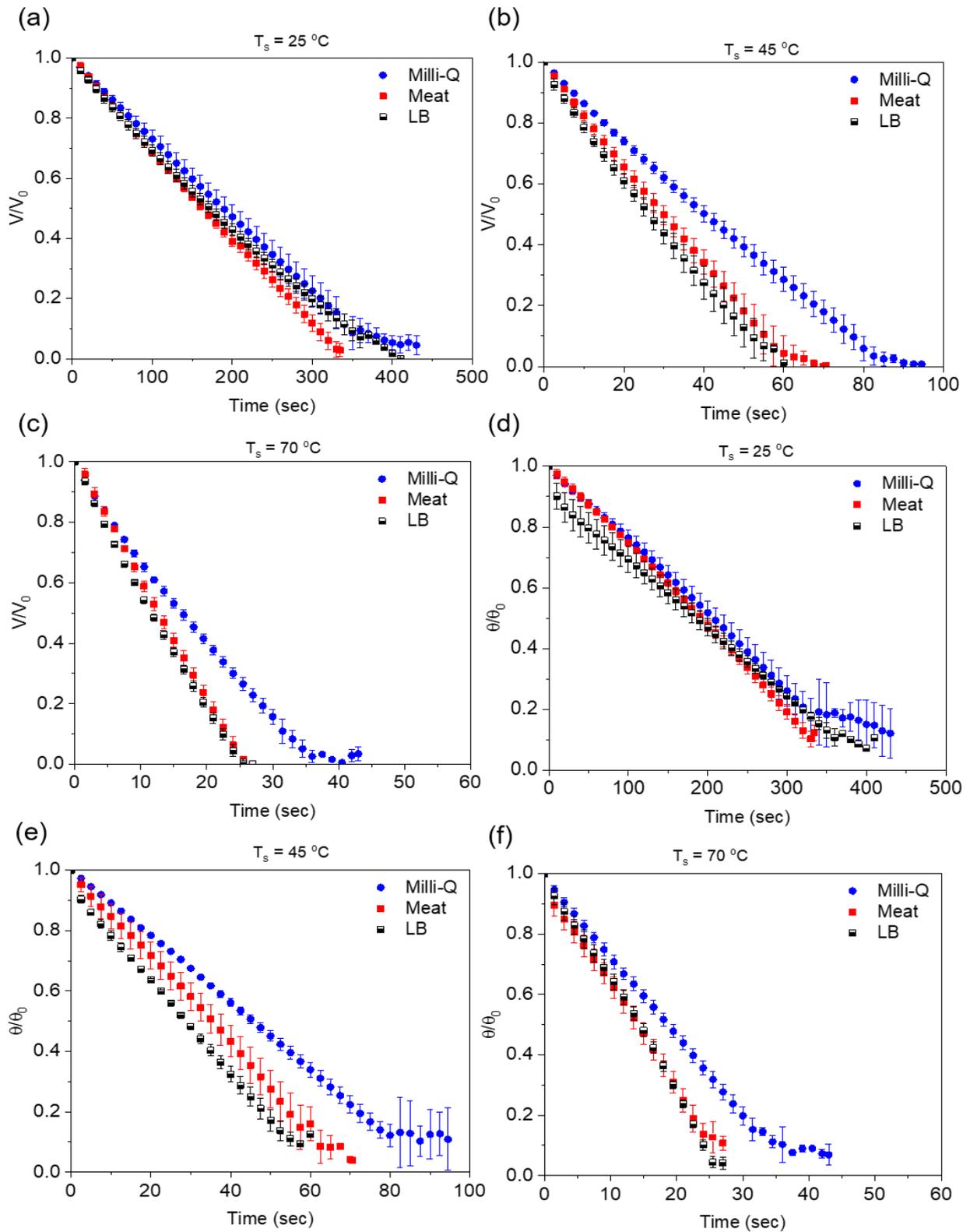

**Figure 3: Normalized droplet parameters versus Time (s) for STM WT in Milli-Q water, LB and Meat mediums for a sessile droplet on glass substrate (a) V/V$_0$ variation at 25ºC (b) V/V$_0$ variation at 45ºC (c) V/V$_0$ variation at 70 ºC (d) θ/θ$_0$ variation at 25ºC (e) θ/θ$_0$ variation at 45ºC temperatures (f) θ/θ$_0$ variation at 70ºC**

Figure 3 depicts the normalized volume variation and contact angle variations for bacteria-laden droplets of Milli-Q water, LB and meat as a medium at 25°C, 45°C, and 70 °C as substrate temperatures. The bacteria-laden Milli-Q water droplet of the same initial volume has $10^8$ CFU/mL STM WT as a solute, but apart from the same concentration of STM WT, the bacteria-laden LB droplet and bacteria-laden meat extract droplet have salt and proteins and only proteins as solutes, respectively. Therefore, the solute content of LB and Meat-based droplets is higher; therefore, their evaporation times are lesser than Milli-Q water-based droplets [37],[38]. The evaporation times for Milli-Q water-based droplets are 400-425 s, 95-105 s, and 38-43 s for 25°C, 45°C, and 70 °C as substrate temperatures, respectively. For 25°C substrate temperature LB based droplets take 370-400 s, and Meat based droplets take 335-360 s. With the increase in substrate temperature, this evaporation time depreciation with respect to Milli-Q water-based fluid gets aggravated, as evident from Figures 3a, 3b, and 3c. For 45°C as substrate temperature, LB-based droplet takes 55-60 s, and Meat droplet took 55-70 s. The LB and Meat-based droplets take approximately 25-30 s at 70°C as substrate temperature.

## 3.2 Infrared Thermography

We performed Infrared (IR) Thermography of the bacteria-laden Milli-Q water droplets to study the temperature profile of the liquid-air interface with respect to time. Figure 4a portrays the temperature scaled images of the droplet on a heated substrate. A particular case of substrate temperature of 70°C at different time frames is shown in the figure. Figure 4b represents the temperature profile of droplet on glass substrate at 25°C.

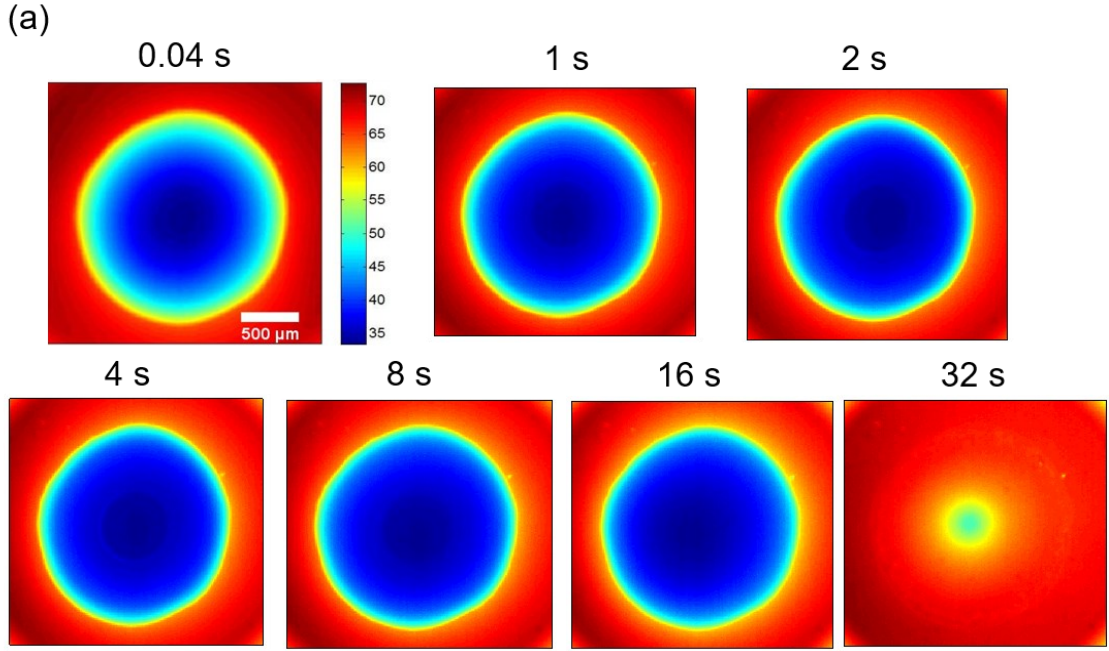

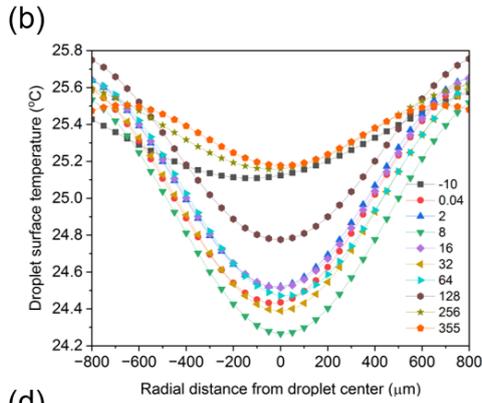
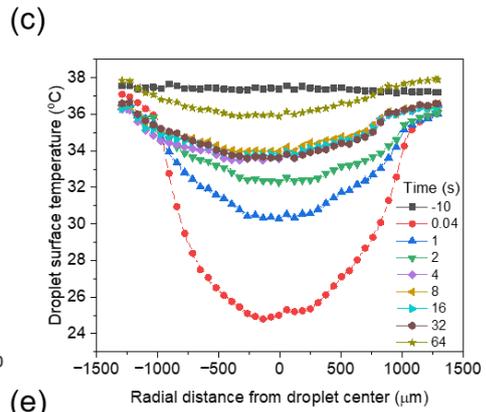
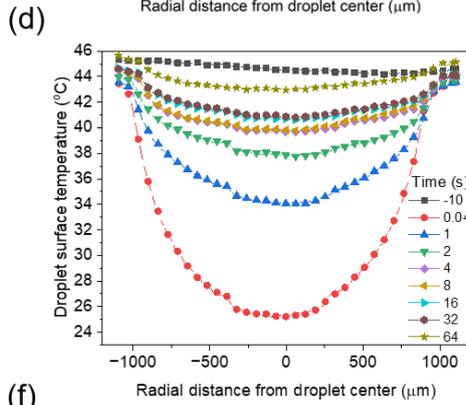
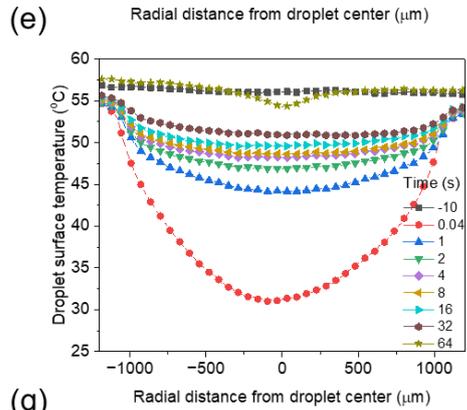
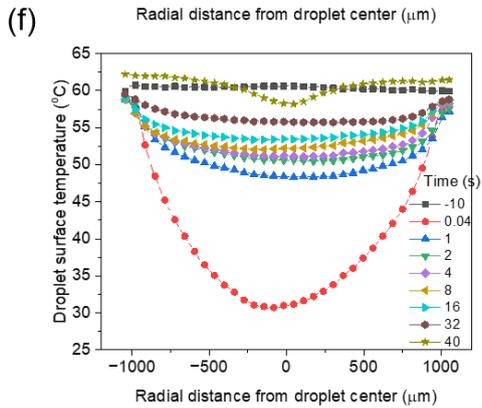
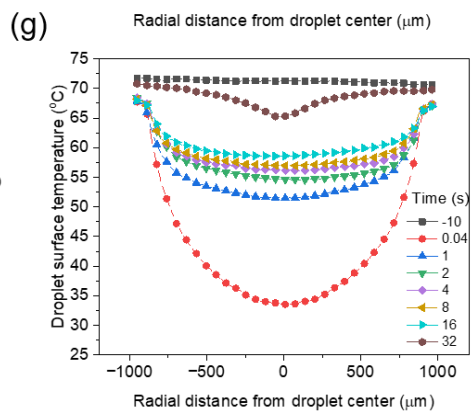

**Figure 4 (a) IR Images of STM WT Milli-Q water droplet from top of the droplet showing the instantaneous isotherms on liquid−gas interface and substrate surface at different time frames for specific case of 70°C substrate temperature. Temperature distribution profiles of droplet surfaces at different times with a spatial resolution of 50 μm for substrate temperature of (b) 25°C (c) 37°C (d) 45°C (e) 55°C (f) 60°C (g) 70°C**

Errors are high since the temperature difference between the substrate temperature (T = 25°C) and initial droplet temperature (T =24.2°C) is significantly less. Therefore, the curve fitting technique shown in Figure 4b is used for plotting. Before being deposited, the droplet is cooler than the surrounding air. Over time, it gradually warms up to match the ambient temperature [30,39]. Experimental results from Figure 4(b-g) show that the temperature at the liquid-air interface at the droplet's top centre is the lowest. This is likely due to the longer heat conduction path and the highest surface tension at that point [40]. The temperature gradient along the liquid-air interface of a droplet on a 25°C substrate is minimal, preventing the induction of Marangoni flow. As a result, capillary flow, driven by higher evaporation at the contact line, becomes the dominant flow mechanism [30],[41],[42].

For Figure 4(c-g), we observed that initially, the substrate temperature was much greater than the droplet top-centre point temperature. The temperature difference was ΔT = 12°C, 20°C, 24°C, 29°C and 37°C at t = 0.04 s for substrate temperatures of 37°C, 45°C, 55°C, 60 °C and 70 °C respectively. The surface tension at the droplet contact line is lower than at the droplet apex point due to the high temperature at the contact line. This temperature-driven difference in surface tension generates the Marangoni flow [32]. In Marangoni flow, the particle/ bacteria are advected towards the droplet apex from the contact line [30],[41]. In addition to the outward capillary flow, a radially inward flow moves along the liquid-air interface from the contact line toward the apex. The two opposite flows create a stagnation region near the contact line, which is responsible for the deposition of the bacteria [43].

The non-dimensional Thermal Marangoni number for the respective substrate temperatures is evaluated using the equation:

$$M_a^T = \frac{-\frac{d\sigma}{dT} R}{\mu \alpha} \Delta T \quad (3.1)$$

Here, dσ/dT is surface tension gradient with respect to temperature in, N m/°C, ΔT (°C) denotes the temperature difference between droplet edge and apex, R is the droplet radius (m), μ is the dynamic viscosity of fluid in, Pa.s and α is the thermal diffusivity in, m²/s. Due to the lower bacterial concentration ($10^8$ CFU/mL) in Milli-Q water, the Thermal Marangoni number is evaluated using the properties of water. Calculations are carried out for all the six the substrate temperatures at the initial time (t= 0.44s). The Thermal Marangoni number varies from $1 \times 10^4$ for 37°C substrate temperature to $5.1 \times 10^4$ at 70°C substrate temperature (refer Supplementary Material). With increase in the substrate temperature, the Thermal Marangoni number increases. The surface tension driven convection along the liquid-air interface is confirmed as the Thermal Marangoni number exceeds the critical values [44].

## 3.3 Flow inside the droplet

We performed live cell imaging for the bacteria-laden Milli-Q water droplet for two substrate temperatures: 1) 25°C and 2) 50°C (see Methods). Figure 5a depicts a schematic of the side view and top view of the droplet on a glass substrate. The region of interest is marked in the schematic. Figure 5b shows velocity contours for a droplet on a substrate temperature of 25 °C at $t/t_f$ =0.2 (t is instantaneous time and $t_f$ is the evaporation time). Similarly Figure 5c displays velocity contours for 50 °C case at $t/t_f$ =0.8.

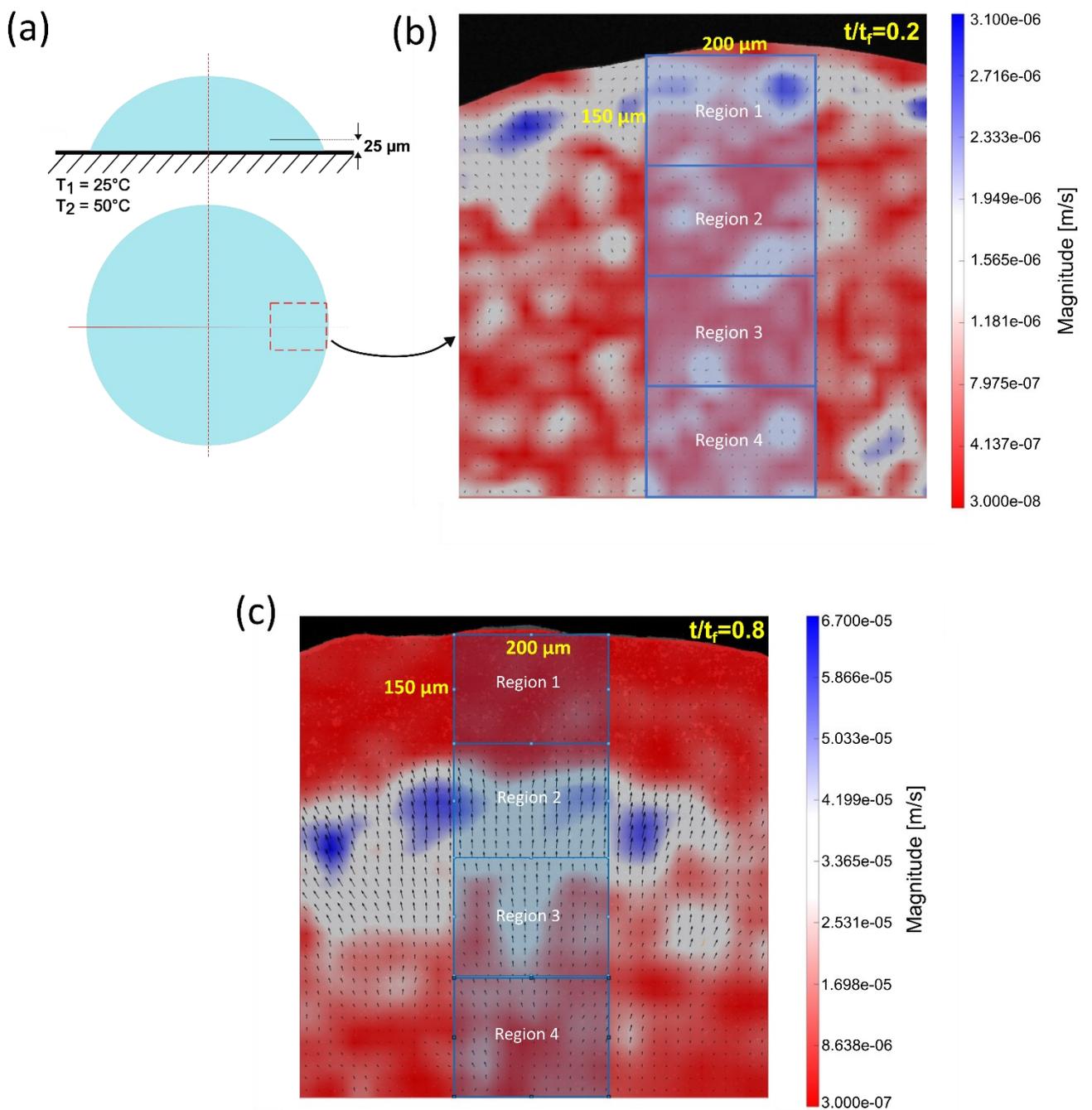

**Figure 5 (a)** Schematic representing confocal live-cell imaging of STM WT Milli-Q water droplet on a glass substrate at 25$^o$C and 50$^o$C temperatures. **(b)** Velocity contours for substrate temperature of 25$^o$C at t/t$_f$ = 0.2 (t is instantaneous time and t$_f$ is the evaporation time) with four regions of 200 μm × 150 μm **(c)** Velocity contours for substrate temperature of 50$^o$C at t/t$_f$ = 0.8 with four regions of 200 μm × 150 μm.

The average velocities at $t/t_f$ =0.2 are the lowest for both the substrate temperature cases across all four regions. In comparison, average velocities at $t/t_f$ =0.8 are highest for both the substrate temperature cases across all four regions (refer Supplementary material Table 1). This is due to the rush-hour effect towards the end of the sessile droplet evaporation [45]. The average velocities of STM WT in a Milli-Q water droplet on a glass slide at 50°C are an order of magnitude higher than those in the same droplet on a glass slide at 25°C. Li et al. measured the velocities near the edge and demonstrated an order-of-magnitude increase with rising substrate temperature [43]. A substrate maintained at 50°C induces a stronger Marangoni flow compared to one at 25°C. Towards the end of evaporation, a Marangoni eddy forms for the case of 50°C substrate temperature (refer to Supplementary Video 2), influencing the bacterial deposition pattern. In contrast, no such eddy formation is observed for the 25°C substrate (refer to Supplementary Video 1). In the previous section, we observed that the non-dimensional Thermal Marangoni number increases with rising substrate temperature. Therefore, the substrate temperature plays a crucial role in governing the internal flow dynamics of the droplet, directly influencing the transport and redistribution of suspended bacteria. This, in turn, determines the final deposition pattern formed upon evaporation.

### 3.4 Micro-characterization of precipitates

Figure 6 displays confocal microscopy images of the dried sessile precipitate samples, captured at maximum intensity. This study was conducted to investigate the impact of substrate temperature on the deposition patterns. Figure 6a depicts the confocal images of dried samples of STM WT in Milli-Q water at the substrate temperatures of 25°C, 37°C, 45°C, 55°C, 60 °C and 70 °C. The sample of 25°C substrate temperature, has a dense ring formation at the periphery of the droplet. The capillary flow dominates at low surface temperatures, causing the formation of ring pattern [46].

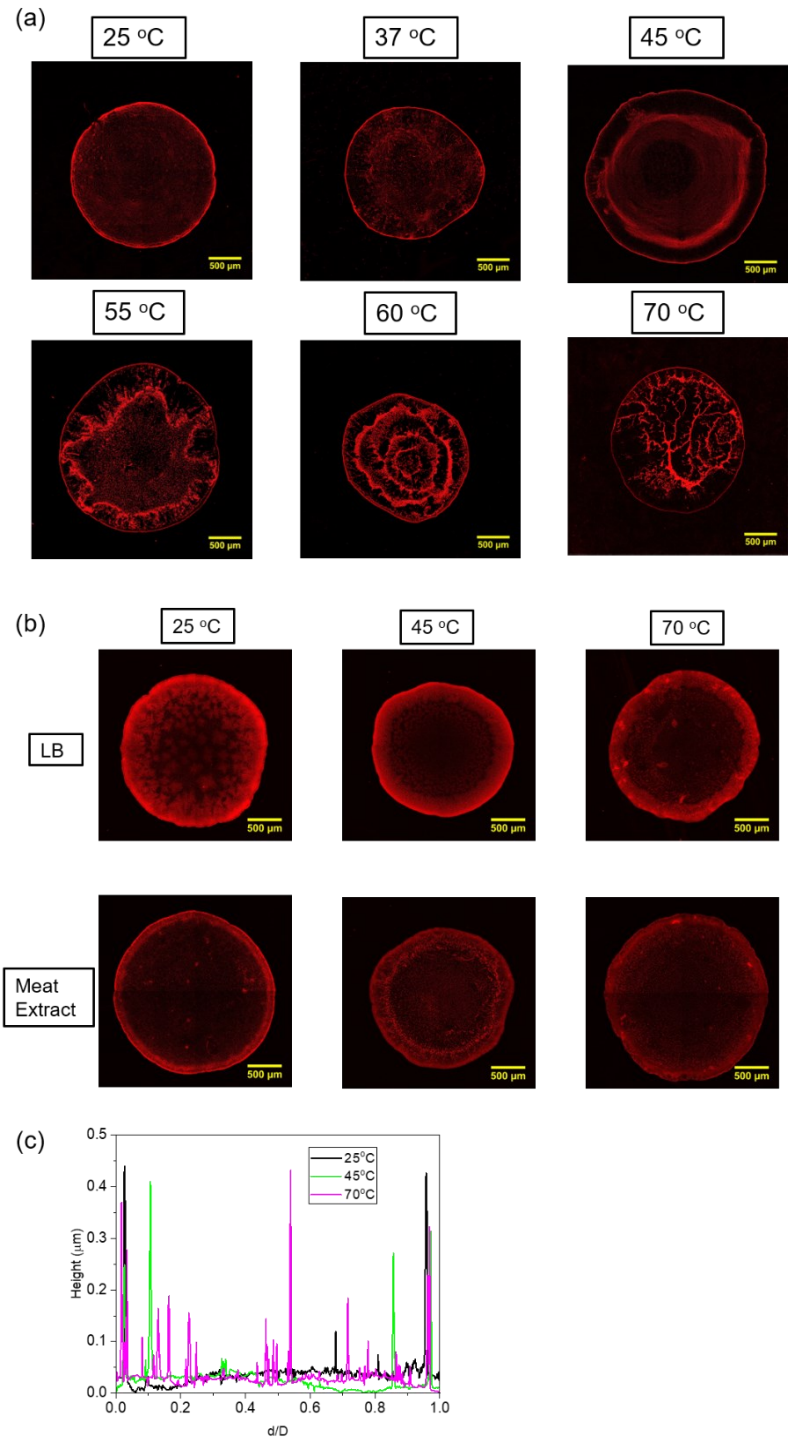

**Figure 6 (a) Confocal microscopy images of precipitates of STM WT in Milli-Q water at 25ºC, 37 ºC, 45 ºC, 55ºC, 60 ºC and 70 ºC substrate temperatures (b) Confocal microscopy images of precipitates of STM WT in LB and Meat extract at 25ºC, 45 ºC, and 70 ºC substrate temperatures (c) Optical profilometry of STM WT in Milli-Q water at 25ºC 45 ºC, and 70 ºC substrate temperatures.**

As the substrate temperature rises, it affects the pattern formation. Therefore, substrate temperature serves as a key parameter influencing the kinetics of pattern development [43]. For the 45°C substrate temperature case, we observe double ring formation. This confirms the presence of Marangoni convection. In a sessile droplet on a substrate maintained at temperatures above ambient, there is a competition between the Marangoni force and viscous stress. Due to the presence of thermal gradients, viscous stress decreases while surface tension stress gradients increase, leading to the receding contact line. Marangoni convection transports bacteria from the contact line toward the droplet apex. During this recirculation process, bacteria are brought onto the substrate, where they are either deposited or carried back to the contact line. Over time, as the contact line recedes, the bacteria accumulate, forming an inner ring [46]. Li et al. explained this phenomenon as a result of the stagnation point shifting due to the movement of the maximum temperature from the contact line to the inner region, caused by higher evaporation at the contact line leading to greater cooling effects [43]. At a substrate temperature of 60°C, a rose-like distribution pattern is observed, whereas at 70°C, the deposition is concentrated at the center. The dead STM WT Milli-Q water fluid forms patterns similar to those of live bacterial fluid (refer to Supplementary Figure), highlighting the significant role of fluid velocities in precipitate pattern formation. This suggests that in an evaporating sessile droplet containing bacteria, convection-driven flow predominantly governs the internal dynamics, rendering bacterial chemotactic movement insignificant [47].

Figure 6b depicts the maximum intensity images of STM WT in LB and Meat mediums at substrate temperatures of 25°C, 45 °C, and 70 °C. The presence of salt and reduced bacterial deposition in the central regions is clearly visible in the LB medium images. For both mediums, temperature-induced variations are most noticeable at the edge or periphery of the precipitate. To gain further clarity, we proceeded with SEM imaging.

Figure 6 c depicts the optical profilometry results of the precipitate of STM WT in Milli-Q water for 25ºC 45 ºC, and 70 ºC substrate temperatures. At 25°C, more bacterial deposition occurs at the edge, which aligns with the confocal image in Figure 6a. The increased deposition at the edge of the precipitate further confirms the dominance of capillary flow and the well-known "coffee-ring"[45].

The two-ring effect is clearly visible in the plot for 45°C, while a central spike appears at the 70°C substrate temperature. Based on the profilometry results, it can be concluded that the edge height is greatest at ambient temperature. Patil et.al. showed via scaling that larger the Marangoni flow velocity the width and height of the deposited ring decreases [30].

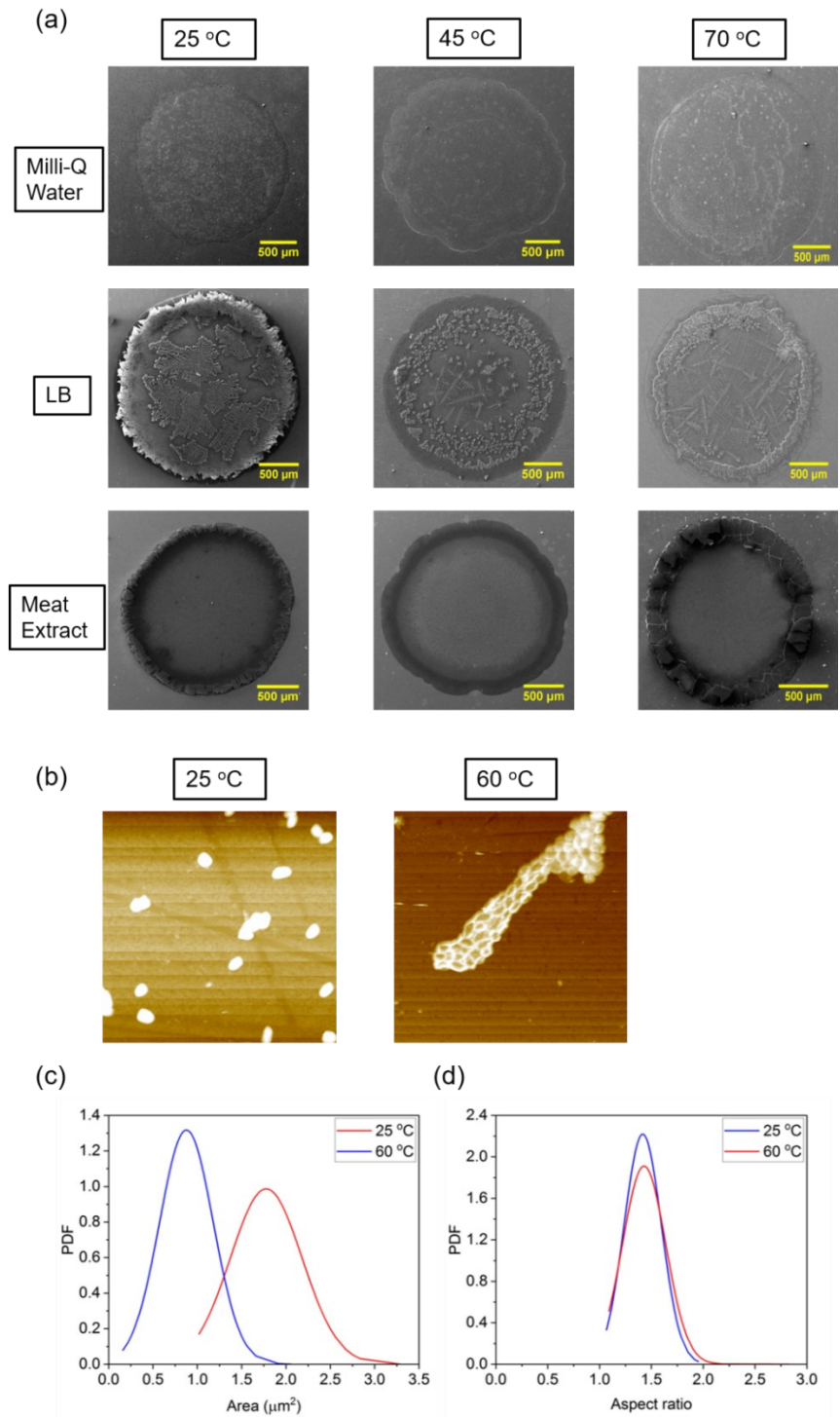

**Figure 7 (a) SEM images of precipitates of STM WT in Milli-Q water, LB and Meat mediums at 25 ºC, 45 ºC, and 70 ºC substrate temperatures (b) AFM images of precipitates of STM WT in Milli-Q water at 25ºC and 60 ºC substrate temperatures and corresponding (c) Areal distribution plot of STM WT (d) Aspect ratio distribution plot of STM WT**

Figure 7a displays the SEM images of precipitate samples of STM WT in Milli-Q water, LB medium, and meat at substrate temperatures of 25°C, 45°C, and 70°C. The images for Milli-Q water closely resemble the corresponding confocal images. In the LB medium samples, dendritic formations are observed at the center, which can be attributed to the presence of salt in the LB medium. LB medium contains tryptone (protein), yeast extract, and NaCl (salt). Binita et al. studied the complex patterns formed by sessile droplets under varying concentrations of salts and proteins [38].

Protein is the primary component of the meat extract-based droplet. The drag force acting on both proteins and bacteria leads to the formation of a thick border at the droplet's periphery, while the central region remains relatively featureless [38]. This thick ring deposition of proteins and bacteria at the droplet's edge is confirmed by Figure 7a and the confocal image in Figure 6b. The stress accumulated within the droplet is relieved through the formation of cracks along the periphery.

Salt alters the interface and electrostatic screening effects, promoting protein aggregation and precipitation. These protein deposits serve as nucleation sites for subsequent salt crystallization. The LB media section in Figure 7a and Figure 6b consistently illustrates this phenomenon. The increased drag force on protein particles causes their early deposition at the droplet's periphery, while salt particles accumulate closer to the center. As the local salt concentration surpasses the saturation limit, it triggers salt crystallization, resulting in dendritic formations at the center.

The variations in the dendritic structures of LB precipitates at substrate temperatures of 25°C, 45°C, and 70°C can be attributed to differences in droplet evaporation rates caused by the increase in temperature [48].

Figure 7b shows AFM images of the central region of Milli-Q water-based STM WT droplets at substrate temperatures of 25°C and 60°C. At 25°C, the bacteria appear scattered in the central region, whereas at 60°C, they form clusters. Surface area and size analyses of the bacteria were performed using the images. Figure 7c shows that the surface area occupied by a single bacterium at 60°C is lesser than at 25°C. The decrease in surface area can be attributed to the higher substrate temperature, which leads to central deposition and cluster formation as a result of increased evaporation rates and evaporative stress. E. Cefali et al. and Abdur et al. suggested that the transformation of bacterial shape from rod to cocoon serves as a protective mechanism against desiccation [49],[50].

However, as shown in Figure 7d, the aspect ratio (the ratio of length to width) of the bacteria remains consistent across both substrate temperatures. Nikola et al. proposed a model explaining how the aspect ratio of rod-shaped bacteria is preserved, regardless of cell size and growth conditions [51].

## 3.5 *Salmonella* can survive at high temperatures on solid, dry conditions, and virulence persists in heat-treated bacteria.

Food contamination by the bacterium *Salmonella* can lead to the gastrointestinal disease salmonellosis, a common foodborne illness [52]. This pathogen is prevalent in various food products, including fruits, vegetables, meat, eggs, and milk [53]. *Salmonella* encounters multiple extreme environmental conditions, such as the acidic pH of food preservatives and temperature fluctuations, which can affect its survival and virulence [54]. Transmission of *Salmonella* occurs primarily through the oral-fecal route via contaminated food and water [55]. In the food industry, this pathogen is exposed to stressors such as high salt concentrations, extreme pH levels, and high or low temperatures [56]. In this study, we first examined whether *Salmonella* can survive in Milli-Q water, LB broth, and meat extract on substrates subjected to different temperature conditions, specifically 25°C, 37°C, 45°C, 55°C, 60°C, and 70°C. Our observations indicated that *Salmonella* could survive at elevated temperatures up to 60°C, but no survival was detected at 70°C (see Figure 8). Next, we hypothesized that exposure to different stress conditions induces varied stress responses, which may impact *Salmonella* virulence. To test this, we infected RAW 264.7 macrophages with heat-treated bacteria. The infection study (see Figure 9) revealed no significant differences in phagocytosis or bacterial proliferation. Interestingly, even at high temperatures, *Salmonella* retained its ability to infect the host at levels comparable to those at optimal temperatures of 25°C and 37°C. Our findings suggest that despite reduced bacterial survival under extreme temperatures, the surviving *Salmonella* cells maintain their infectivity characteristics

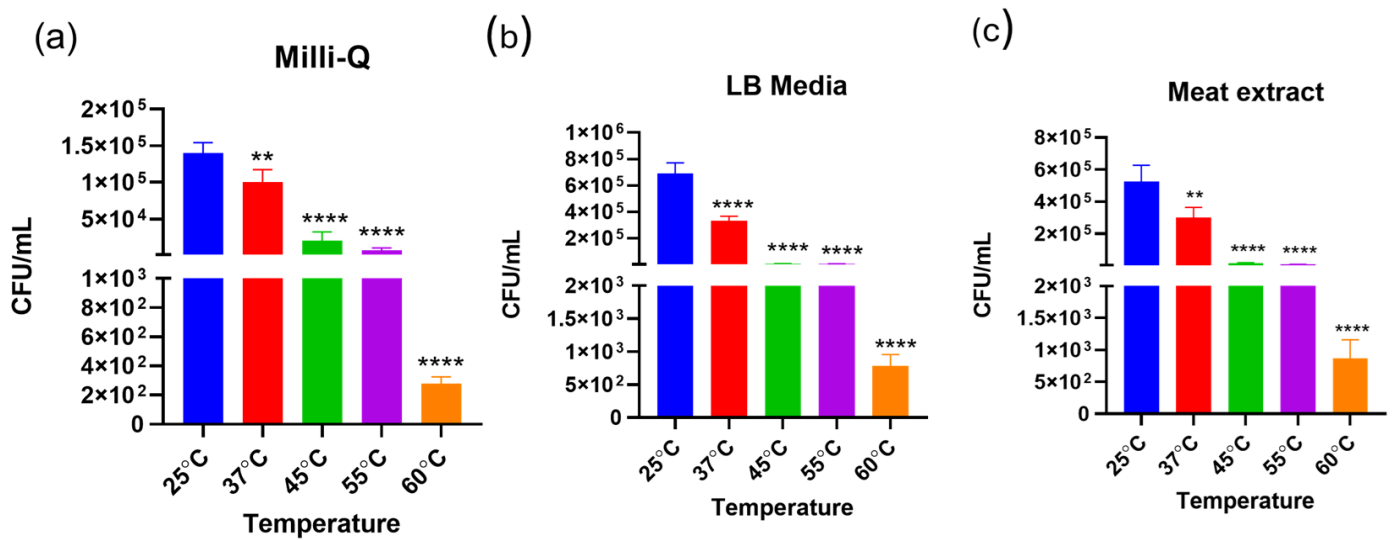

**Figure 8 Survival assay of *Salmonella* in (A) Milli-Q, (B) LB media, and (C) Meat extract at different ranges of temperature. Student's t-test was used to analyse the data. All data are represented as mean ± SD from three independent experiments.; p values * ** *< 0.0001, * ** < 0.01, * <0.05.**

Our study indicates that *Salmonella* can survive at elevated temperatures, though its survivability decreases as temperatures rise. Pasteurization, a widely used method to eliminate harmful bacteria in food and beverages, typically employs moderate temperatures ranging from 60°C to 80°C [57]. Consistent with this, our findings show that *Salmonella* survival diminishes at higher temperatures. However, more importantly, the surviving bacteria at 45°C, 55°C, and 60°C retained their ability to infect host cells at levels comparable to those observed at 25°C and 37°C.

However, our study does not reveal precisely how *Salmonella* regulates their virulence at high temperatures. Future studies on regulating genes at different temperatures will provide further mechanistic depth to our observations understanding how *Salmonella* responds to heat stress while maintaining its infectivity. Understanding these mechanisms may contribute to improved food safety strategies and thermal processing techniques [54].

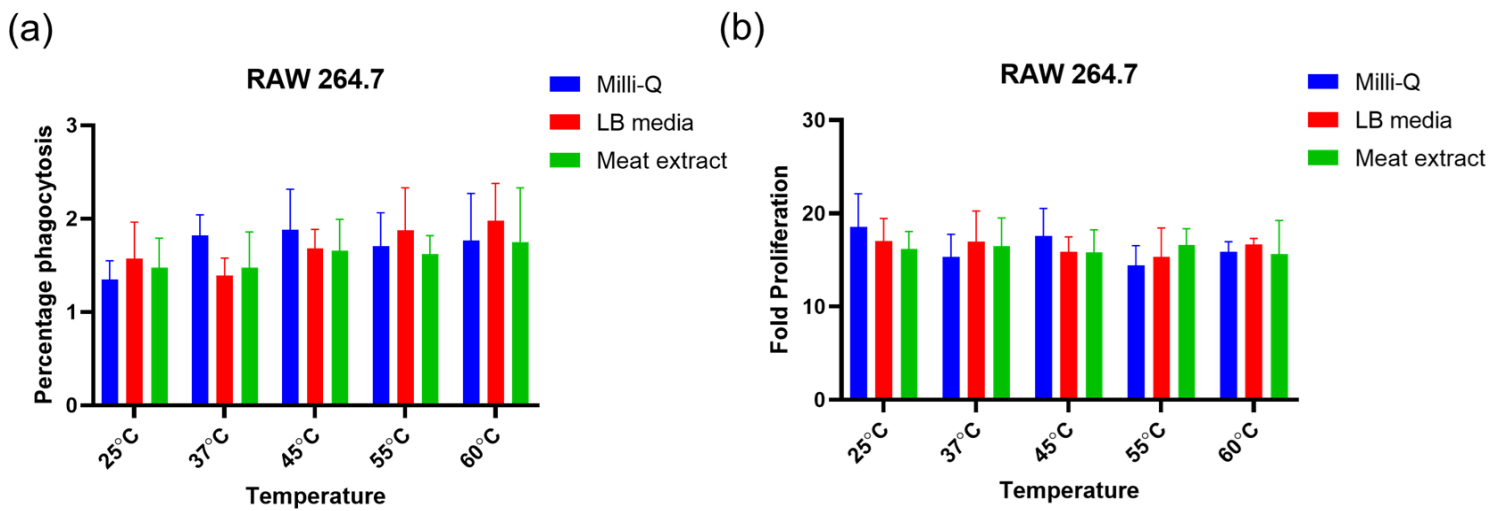

**Figure 9 Intracellular survival assay in RAW264.7 macrophage cell lines with heat-treated bacteria (D) Percentage phagocytosis, (E) Fold proliferation. Represented data of N=3, n=3. Student's t-test was used to analyse the data Two-way ANOVA with Tukey's post-hoc test was used to analyse the grouped data. All data are represented as mean ± SD from three independent experiments.; p values * ** *< 0.0001, * ** < 0.01, * <0.05**.

# 4. Conclusions

Critical insights into the interplay between droplet desiccation, mass transport phenomena, and bacterial survival, emphasizing the significant role of substrate temperature and properties, are provided by our study. At ambient temperatures on hydrophilic surfaces, capillary flow governs the evaporation process yielding the classic coffee-ring effect in bacterial deposition [30],[32],[45]. However, at elevated substrate temperatures, the thermal gradients induce surface tension variations, generating inward Marangoni flows. For instance, at the 50°C case, the radial velocity increases by an order of magnitude relative to the 25°C case, with pronounced Marangoni eddies altering the final deposition patterns [30][32][41][43][44].

Our experiments reveal that Salmonella deposition patterns and survivability are strongly influenced by substrate temperature and the composition of the base fluid. Salmonella's deposition pattern in Milli-Q water shifts from a single edge ring at 25°C case to double rings at 45°C case and central deposition at 70°C case. In contrast, droplets prepared with LB media or meat extract display additional complexity, where proteins and salts interact with bacteria to form unique deposition structures [38]. These findings not only extend classical colloid and interface science concepts but also highlight how thermal gradients and solute interactions drive complex behavior in biologically active systems [40][58].

As the substrate temperature increases, the bacterium reduces its surface area while maintaining its aspect ratio, likely as a defense mechanism against evaporative stress and evaporation [51]. The pathogen's resilience and raising concerns for food safety are underscored, as despite exposure to high temperatures, the surviving Salmonella remains as infectious as those from lower-temperature cases (25°C and 37°C). This observation suggests that traditional thermal treatments, such as pasteurization (typically 60°C–80°C) [57], may require further optimization to inactivate robust bacterial populations effectively.

This work combines colloidal dynamics with microbiological assessments to show how thermal and interfacial forces affect bacterial deposition and survival. Our study demonstrates that thermal Marangoni flows significantly impact bacterial behavior—a step forward compared to studies that focused only on capillary effects and reveals that even under high-temperature stress, surviving Salmonella remains infective. Future research should investigate the genetic and molecular pathways that allow Salmonella to regulate virulence under heat stress and include comparative studies across different food matrices and conditions to refine decontamination strategies and improve food safety protocols. Ultimately, this work bridges colloid science with microbiology, and a novel perspective on pathogen survival mechanisms is provided that can inform improved industrial practices and risk management in food processing.

## **Supplementary Material**

Figure S1: Confocal microscopy images of precipitates of dead STM WT in Milli-Q water at 25°C, 37 °C, 45°C, 55°C, 60°C and 70°C substrate temperatures.

Table T1: Average velocities of STM WT RFP in Milli-Q water in μm/s at $t/t_f$ = 0.2, 0.4, 0.6 and 0.8 (t is instantaneous time and $t_f$ is the evaporation time) for regions 1, 2, 3 and 4 of 200 μm × 150 μm at substrate temperature (a) 25°C (b) 50°C.

Section S1: Steady-state substrate temperature

Section S2: p values and Statistical analysis

Video V1: Confocal live cell video (speed of 13×) of STM WT RFP at 25°C substrate temperature.

Video V2: Confocal live cell video (speed of 2×) of STM WT RFP at 50°C substrate temperature.

# Acknowledgements

SB acknowledges financial support from the Science and Engineering Research Board (SERB), Pratt & Whitney Chair Professorship and SERB-SUPRA (Scientific and Useful Profound Research Advancement) project number SERB/F/10572/2021-2022. This work was supported by the DAE SRC fellowship (DAE00195) and DBT-IISc partnership umbrella program for advanced research in biological sciences and Bioengineering to DC. Infrastructure support from ICMR (Centre for Advanced Study in Molecular Medicine), DST (FIST), and UGC (special assistance) is highly acknowledged along with ASTRA- Chair fellowship, TATA Innovation grant, and DBT-IOE partnership grant to DC. ANA gratefully acknowledges the Prime Minister's Research Fellows (PMRF) fellowship scheme facilitated by the Ministry of Higher Education (MHRD), Government of India. ANA acknowledges the JEOL- SEM IT 300 facility at Advanced Facility for Microscopy and Microanalysis (AFMM), IISc, AFM Facility at Department of BSSE (Biosystems Science and Engineering), Andor-Dragonfly facility at Biological Science Central Confocal Imaging Facility, and the MCB (Microbiology and Cell Biology) Confocal Imaging Facility, IISc. Jason Joy Poopady is acknowledged for his contribution in the confocal flow experiment part. AS sincerely acknowledges the UGC fellowship for his financial support. The funders had no role in study design, data collection, and analysis, publication decisions, or manuscript preparation.

# Author Contribution

SB, DC, and ANA have conceptualized the study. ANA and AS have contributed to the experiment designing, visualization, methodology, investigation, formal analysis, literature survey, validation, writing (original draft), reviewing, and editing of the manuscript. KKD performed the experiment, methodology, visualization, investigation, literature survey, and manuscript editing along with AS and ANA. SB and DC have contributed to the funding acquisition, project administration, and the overall work supervision. All authors took part in writing, reviewing, and editing the manuscript and approved the final version.

## Abbreviations

SB – Prof. Saptarshi Basu

DC – Prof. Dipshikha Chakravortty

ANA- Amey Nitin Agharkar

AS – Anmol Singh

KKD- Dr. Kush Kumar Dewangan

## Conflict of Interest

The authors have declared that no conflict of interest exists.

## Data Availability Statement

The data that supports the findings of this study are available within the article and its supplementary material.

**Address correspondence to**:

Prof. Saptarshi Basu, 406, Department of Mechanical Engineering, Indian Institute of Science (IISc), Bangalore 560012, India, Telephone number: +91 80 22933367, Email: sbasu@iisc.ac.in

Prof. Dipshikha Chakravortty, SA-10, Department of Microbiology and Cell Biology, Indian Institute of Science (IISc), Bangalore 560012, India, Telephone number: +91 80 22932842, Email: dipa@iisc.ac.in